\documentclass[usenatbib]{mnras}
\usepackage{lineno}
\usepackage{natbib}
\usepackage{graphicx}
\usepackage{amssymb}
\usepackage{color}
\usepackage{upgreek}
\usepackage{amsmath}
\usepackage{soul}

\newcommand{\ptrsa}{RSPTA}
\newcommand{\cursci}{Curr. Sci.}

\newcommand{\ass}{Ap\&SS}

\author[Perera et al.]{B.~B.~P.~Perera$^{1}$, B.~W.~Stappers$^{1}$,  S.~Babak$^{2,3}$, M.~J.~Keith$^{1}$, J.~Antoniadis$^4$, 
\newauthor C.~G.~Bassa$^{5}$, R.~N.~Caballero$^4$, D.~J.~Champion$^4$, I.~Cognard$^{6,7}$, G.~Desvignes$^4$, 
\newauthor E.~Graikou$^{4}$, L.~Guillemot$^{6,7}$, G.~H.~Janssen$^{5,8}$, R.~Karuppusamy$^4$, M.~Kramer$^{4,1}$, 
\newauthor P.~Lazarus$^4$, L.~Lentati$^{9}$, K.~Liu$^4$, A.~G.~Lyne$^{1}$, J.~W.~McKee$^{1,4}$, S.~Os{\l}owski$^{10,4,11}$, 
\newauthor D.~Perrodin$^{12}$, S.~A.~Sanidas$^{13,1}$, A.~Sesana$^{14}$, G.~Shaifullah$^{11,4,5}$, G.~Theureau$^{6,7,15}$, 
\newauthor J.~P.~W.~Verbiest$^{11,4}$, S.~R.~Taylor$^{16}$  
\\ $^1$ Jodrell Bank Centre for Astrophysics, School of Physics and Astronomy, The University of Manchester, Manchester M13 9PL, UK
\\ $^2$ APC, Univ. Paris Diderot, CNRC/IN2P3, CEA/lrfu, Obs. de Paris, Sorbonne Paris Cite, France
\\ $^3$ Moscow Institute of Physics and Technology, Dolgoprudny, Moscow region, Russia
\\ $^4$ Max-Planck-Institut f$\ddot u$r Radioastronomie, Auf dem H$\ddot u$gel 69, D-53121 Bonn, Germany
\\ $^5$ ASTRON, the Netherlands Institute for Radio Astronomy, Postbus 2, 7990 AA, Dwingeloo, the Netherlands
\\ $^6$ Laboratoire de Physique et Chimie de l'Environnement et de l'Espace LPC2E CNRS-Universit\'{e} d'Orl\'{e}ans, F-45071 Orl\'{e}ans, France
\\ $^7$ Station de radioastronomie de Nan\c{c}ay, Observatoire de CNRS/INSU, F-18330 Nan\c{c}ay, France
\\ $^8$ Department of Astrophysics/IMAPP, Radboud University, P.O. Box 9010, 6500 GL Nijmegen, The Netherlands
\\ $^9$ Astrophysics Group, Cavendish Laboratory, JJ Thomson Avenue, Cambridge CB3 0HE, UK
\\ $^{10}$ Centre for Astrophysics and Supercomputing, Swinburne University of Technology, Post Box 218 Hawthorn, VIC 3122, Australia
\\ $^{11}$ Fakult$\ddot{a}$t f$\ddot{u}$r Physik, Universit$\ddot{a}$t Bielefeld, Postfach 100131, D-33501Bielefeld, Germany 
\\ $^{12}$ INAF-Osservatorio Astronomico di Cagliari, via della Scienza 5, I-09047 Selargius (CA), Italy
\\ $^{13}$ Anton Pannekoek Institute for Astronomy, University of Amsterdam, Science Park 904, NL-1098 XH Amsterdam, the Netherlands
\\ $^{14}$ School of Physics and Astronomy, University of Birmingham, Edgbaston, Birmingham B15 2TT, UK
\\ $^{15}$ Laboratoire Univers et Th\'{e}ories LUTh, Observatoire de Paris, CNRS/INSU, Universit\'{e} Paris Diderot, 5 place Jules Janssen, 
\\F-92190 Meudon, France
\\ $^{16}$ Jet Propulsion Laboratory, California Institute of Technology, Pasadena, CA 91109, USA
}

\title[GWs limits from J1713$+$0747 high-cadence data]{Improving timing sensitivity in the microhertz frequency regime: limits from PSR J1713$+$0747 on gravitational waves produced by super-massive black-hole binaries}

\begin{document}

\maketitle

\begin{abstract}
We search for continuous gravitational waves (CGWs) produced by individual super-massive black-hole binaries (SMBHBs) in circular orbits using high-cadence timing observations of PSR J1713$+$0747. We observe this millisecond pulsar using the telescopes in the European Pulsar Timing Array (EPTA) with an average cadence of approximately 1.6 days over the period between April 2011 and July 2015, including an approximately daily average between February 2013 and April 2014. The high-cadence observations are used to improve the pulsar timing sensitivity across the GW frequency range of $0.008-5$~$\upmu$Hz. We use two algorithms in the analysis, including a spectral fitting method and a Bayesian approach. For an independent comparison, we also use a previously published Bayesian algorithm. We find that the Bayesian approaches provide optimal results and the timing observations of the pulsar place a 95 per cent upper limit on the sky-averaged strain amplitude of CGWs to be $\lesssim3.5\times10^{-13}$ at a reference frequency of 1~$\upmu$Hz. We also find a 95 per cent upper limit on the sky-averaged strain amplitude of low-frequency CGWs to be $\lesssim1.4\times10^{-14}$ at a reference frequency of 20~nHz. 
\end{abstract}


\section{Introduction}

Millisecond pulsars (MSPs) are old neutron stars that are spun up to periods $\lesssim 30$~ms during a mass accretion phase in binary systems through a so-called recycling process \citep{rs82,acrs82}. They are among the most stable known rotators in the universe and the long-term stability, on a time scale of $>$10~yr, of the rotation of some of them is comparable to that of an atomic clock \citep[e.g.][]{pt96,hcm+12}. Due to their high rotational stability, the timing measurements of MSPs can be obtained with a significant accuracy \citep[e.g.][]{ktr94}. Therefore, pulsars have been identified as excellent tools for searching for gravitational waves (GWs) \citep[see][]{det79,hd83,jhlm05}.  

Pulsar timing is sensitive to low-frequency GWs, from nHz to $\upmu$Hz \citep[see][]{det79,hd83,jhlm05}, the natural frequency range of inspiralling SMBHBs. Forming as a consequence of galaxy mergers, SMBHBs are expected to form frequently across cosmic history \citep[e.g.][]{vhm03}. The superposition of GW signals from this cosmic population results in a stochastic gravitational wave background \citep{rr95a, jb03, svc08}, although particularly nearby massive systems are likely to result in individually resolvable signals rising above the background level \citep{svv09}. The lower frequency limit of GWs that pulsar timing is sensitive to is $1/T_{\rm obs}$, where $T_{\rm obs}$ is the total observation span, indicating that a longer time-baseline of data helps to probe the low-frequency GWs. The upper frequency limit depends on the data sampling pattern since the observations are not evenly sampled. Most of the previous pulsar-based GW studies have used the traditional averaged Nyquist frequency $N/(2T_{\rm obs})$ as the upper limit, where $N$ is the number of observations \citep[e.g.][]{yhj+10,yss+14,zhw+14,abb+14,bps+16}. To be consistent with previous studies and compare the results, we restrict our search analysis to the same frequency range. We note that the upper frequency limit for an unevenly sampled data set can be much larger than the averaged Nyquist frequency \citep[see][]{eb99,van17}, but determining the sensitivity in this frequency range is complicated. In order to probe these low-frequency GWs, high-precision timing of MSPs with sub micro-second timing accuracy is essential. Pulsar Timing Arrays (PTAs) --  the European PTA \citep[EPTA;][]{dcl+16}, the Parkes PTA \citep[PPTA;][]{rhc+16}, and the North American Nanohertz Observatory for Gravitational Waves \citep[NANOGrav:][]{abb+15} -- provide a unique way to obtain such a high timing accuracy by observing a collection of highly stable MSPs with good cadence over a long-term period. The International PTA (IPTA), a collaboration of the three individual PTAs, observes about 50 MSPs in total with a time baseline range of $\sim(4.5-27)$~yr and an approximately weekly to monthly cadence, including short-term daily observation campaigns \citep[see][]{vlh+16}.

The limits on the amplitude, or the strain, of GWs produced by individual SMBHBs have been previously investigated through pulsar timing in several studies \citep[e.g.][]{lb01,jllw04}. \citet{bps+16} used 41 MSPs in the first EPTA data release \citep{dcl+16} and estimated the sky-averaged strain amplitude of CGWs to be in the range of $(0.6 - 1.5)\times10^{-14}$ at a frequency of $5-7$~nHz. By using 20 MSPs in the first PPTA data release \citep{mhb+13}, \citet{zhw+14} estimated the upper limit on the strain amplitude of CGWs to be $1.7\times10^{-14}$ at a frequency of 10~nHz. \citet{abb+14} used the observations of 17 MSPs reported in \citet{dfg+13} and placed the upper limit on the strain of CGWs to be $3\times10^{-14}$ at a frequency of 10~nHz.

Using a spectral fitting method, \citet{yhj+10} constrained the upper limit on the strain amplitude of CGWs to be about $\sim 1\times10^{-13}$ at a frequency of $\sim 9$~nHz based on 18 MSPs observed by the PPTA. Following the same method, \citet{yss+14} estimated the upper limit on the strain amplitude of CGWs based on high-cadence observations of PSR B1937$+$21. They found that the timing of this pulsar constrains an upper limit on the strain amplitude of $\lesssim1.5\times10^{-11}$ and $\lesssim5\times10^{-14}$ at $10^{-7}$~Hz for random and optimal source location and polarisation of individual GW sources, respectively. In the timing analysis, they noticed an unmodelled periodic noise in the timing residuals (i.e. the difference between the observed and model-predicted time of arrivals of pulses) of PSR B1937$+$21 with an amplitude of 150~ns at a frequency of 3.4~yr$^{-1}$. \citet{yss+14} subtracted this periodic signal by fitting a sinusoid and then used the whitened residuals in the GW search analysis. It has been previously reported in other studies that this pulsar exhibits a high level of red timing noise \citep[see][]{cll+16,lsc+16} and a significant dispersion measure (DM) variation \citep[e.g. ][]{ktr94,mhb+13,abb+15,dcl+16}, where DM accounts for the frequency-dependent time delay of the radio pulses due to electrons in the interstellar medium along the line-of-sight. Therefore, the noise and the DM variation of the pulsar may have perhaps generated this particular periodic signal that was seen in the data. \citet{bps+16} excluded PSR B1937$+$21 in their GW analysis due to its complicated high-level of timing noise. We note that PSR J1713$+$0747 exhibits a much lower level of timing noise and DM variation across our observation time span \citep[see][]{dcl+16,abb+15}. We thus emphasise that PSR J1713$+$0747 is a better choice compared to PSR B1937$+$21 for these types of single-pulsar timing-based GW search analyses. We also note that single-pulsar GW search analyses cannot conclusively detect GWs, rather produce upper limits on strain amplitudes.

PSR J1713$+$0747 is one of the most precisely timed pulsars by PTAs \citep{dcl+16,rhc+16,abb+15}. It is regularly observed by all the telescopes involved in PTAs, providing a timing stability of $\sim100$~ns based on the recent NANOGrav data release which contains over 9 years of observations \citep{abb+15}. \citet{zsd+15} reported a timing stability of $\sim 92$~ns based on 21 years of observations of the pulsar. \citet{sc12} investigated the stability of the pulsar and reported that single pulses show an rms phase fluctuation (the so-called pulse jitter) of $\sim40$~ns, which limits the timing precision. Recently, \citet{lbj+16} analysed the single pulses of the pulsar in detail with data collected by the Large European Array for Pulsars \citep[LEAP --][]{bjk+16} project and confirmed that the pulsar shows two modes of systematic sub-pulse drifting \citep{es03}. The IPTA undertook a 24-hr continuous global observation campaign of this pulsar using nine telescopes around the world \citep{dlc+14}. \citet{dec+16} showed that the data set in this campaign is sensitive to CGWs with a frequency between $0.01-1$~mHz. Following \citet{yhj+10}, \citet{dec+16} estimated the upper limit on the strain amplitude of CGWs produced by the sources located in the direction of the pulsar to be $\sim 10^{-11}$ at a frequency $0.01$~mHz.

The high-cadence observations are important to understand the noise, and the GW signal if it is present, in pulsar timing data. Since the observations are unevenly sampled, decreasing the relative cadence of the data will increase the noise level across the power spectrum of the timing residuals. The paper is organised as follows: in Section~\ref{obs}, we present our observations and data processing. We describe the timing analysis of the pulsar in Section~\ref{timing} and present our timing measurements and residuals. We then present our analysis carried out to obtain the upper limit on the strain amplitude of CGWs in Section~\ref{spectral} and \ref{bayes} using a spectral fitting and a Bayesian approach, respectively. For an independent comparison, we use the algorithm described in \citet{bps+16} on our data set and present the results in Section~\ref{babak}. We apply our limits on SMBHB candidates in Section~\ref{application} and investigate any possibilities of detecting them. Finally in Section~\ref{dis}, we summaries our results.

\section{Observations and data processing}
\label{obs}

PSR J1713$+$0747 was observed using the Lovell Telescope (LT) at the Jodrell Bank Observatory between 2011 April and 2015 July with an average cadence of about $4.5$ days. Between 2013 February and 2014 April, the pulsar was observed more frequently with an average cadence of approximately every two days. All the LT observations were carried out at L-band (with a centre frequency of 1532~MHz) and the data were recorded using the `ROACH' pulsar backend \citep[see][]{bjk+16}. In addition to these high-cadence LT observations, we included the observations of the pulsar obtained from three other telescopes -- Effelsberg Telescope (EFF) in Germany, Nan\c{c}ay Radio Telescope (NRT) in France, and Westerbork Synthesis Radio Telescope (WSRT) in the Netherlands -- in the analysis to further improve the cadence. The EFF observations were recorded using the `PSRIX' backend \citep[see][]{lkg+16} and carried out at two centre frequencies, 1347 and 2627~MHz. Both single- and multi-beam L-band receivers were used at EFF alternatively throughout our observation time span. The NRT observations were made at centre frequencies of 1484, 1524 and 2539~MHz using the `NUPPI' backend \citep[see][]{ct06,ctg+13}. The WSRT observations in this analysis were recorded using the `PuMa~II' backend \citep[see][]{ksv08} and observed at three centre frequencies 350, 1380, and 2200~MHz. All these backends used DSPSR to perform coherent dedispersion and folding \citep{vb11}. We note that these observations were obtained using newer backends and are different from the data sets reported in the EPTA data release \citep{dcl+16}. To be consistent with all observations, we selected the beginning and the end time of each data set to be equal to that of the LT observations, but we  note that the NRT observations started four months after the beginning of the LT observations. The details of the bandwidths, data span, and the number of observations in each data set is given in Table~\ref{info}. Combining the data from all four telescopes, 952 observation epochs were included in our analysis with a resulting average cadence of about 1.6 days. We also note that the combination of data from other telescopes improved the average cadence of the pulsar to approximately daily within the period between 2013 February and 2014 April, including 420 TOAs in total, in which it was monitored more frequently using the LT.

\begin{table*}
\begin{center}
\caption{
Details of observations used in the analysis. Two $1.347$~GHz Effelsberg data sets have been included in the analysis; the data obtained using the multi-beam receiver (denoted as `$^{\rm m}$') and the single-beam receiver (denoted as `$^{\rm s}$'). The observation length varies and typical values for data sets are given. In total, 952 observation epochs were included in the analysis, resulting in an average cadence of about 1.6 days. The last column shows EFAC and $\rm \log_{10} [EQUAD(s)]$ determined for each data set in the timing model given in Table~\ref{solution}. 
}
\label{info}
\begin{tabular}{lccccccccc}
\hline
\multicolumn{1}{c}{Data set} &
\multicolumn{1}{c}{Backend} &
\multicolumn{1}{c}{Centre Freq} &
\multicolumn{1}{c}{Bandwidth} &
\multicolumn{1}{c}{Channel size} &
\multicolumn{1}{c}{Phase} &
\multicolumn{1}{c}{Obs. length} &
\multicolumn{1}{c}{Data span} &
\multicolumn{1}{c}{No. of} &
\multicolumn{1}{c}{E$_{\rm f}$,} \\
\multicolumn{1}{c}{ } &
\multicolumn{1}{c}{ } &
\multicolumn{1}{c}{(MHz)} &
\multicolumn{1}{c}{(MHz)} &
\multicolumn{1}{c}{(MHz)} &
\multicolumn{1}{c}{bins} &
\multicolumn{1}{c}{(min)} &
\multicolumn{1}{c}{ } &
\multicolumn{1}{c}{ TOAs} &
\multicolumn{1}{c}{$\rm \log_{10}[E_{\rm q}(s)]$} \\
\hline
\hline
LT & ROACH & 1532 & 400 & 1 & 2048$^a$ & 10/30/60 & 4/2011 -- 7/2015 & 336 & 1.00, $-6.81$ \\
\hline
NRT & NUPPI & 1484 & 512 & 4 & 2048 & 42 & 8/2011 -- 7/2015 & 193 & 1.23, $-6.98$ \\
    &     & 1524 & 512 & 16 & 2048 & 45 & 1/2013 -- 5/2015 & 20 & 1.11, $-7.31$ \\
    &     & 2539 & 512 & 4 & 2048 & 52 & 8/2011 -- 12/2014 & 37 & 1.34, $-6.57$ \\ 
\hline
EFF & PSRIX & 1347$^{\rm m}$ & 200 & 1.56 & 1024 & 28 & 4/2011 -- 5/2015 & 56 & 1.25, $-6.59$ \\
   &  & 1347$^{\rm s}$ & 200 & 1.56 & 1024 & 28 & 5/2011 -- 5/2013 & 27 & 1.42, $-6.83$ \\
   &       & 2627 & 200 & 1.56 & 1024 & 30 & 4/2011 -- 6/2015 & 41 & 1.06, $-6.22$ \\
\hline
WSRT & PuMaII & 350 & 70 & 0.156 & 256 & 25/40/60 & 4/2011 -- 7/2015 & 87 & 1.11, $-9.80$ \\
     &      & 1380 & 160 & 0.312 & 512 & 25/45 & 4/2011 -- 7/2015 & 123 & 0.97, $-6.47$ \\
     &      & 2200 & 160 & 0.312 & 512 & 25/40 & 4/2011 -- 12/2014 & 32 & 0.67, $-7.09$ \\
\hline
\end{tabular}
\begin{tabular}{l}
\multicolumn{1}{l}{} \\
$^a$ Note that there were only 512 pulse phase bins in the early LT observations (i.e. before 2011 July -- MJD 55754). Therefore, we fit \\for an additional `JUMP' in the timing analysis to correct for the time delay of these TOAs.
\end{tabular}
\end{center}
\end{table*}

We processed the data using the pulsar analysis software package PSRCHIVE\footnote{http://psrchive.sourceforge.net/} \citep{hvm04,vdo12}. PSR J1713$+$0747 is well known to show a frequency-dependent pulse profile shape variation \citep[see][]{dlc+14,abb+15,zsd+15}, even within a frequency range of a wide-band receiver ($\sim$400~MHz). This may add an extra uncertainty in the measured time-of-arrival (TOA) if it is obtained through the standard bandwidth-averaged TOA generating method \citep{tay92}. We also note that the scintillation of the pulsar may lead to significant changes in the bandwidth-averaged pulse profile shape due to attenuation of the intensity of the different parts of the frequency band in different observations. Therefore, we use broad-band TOA measurement techniques \citep[see][]{pdr14,ldc+14} on LT and NRT observations. Each epoch was first folded  for the entire observation duration using a previously published timing model of the pulsar \citep{dcl+16}, while keeping the full frequency resolution across the bandwidth. This gives better signal-to-noise (S/N) in frequency-channel-dependent pulse profiles. We then sum neighbouring frequency channels together to form eight equal width sub-bands to further improve the S/N. We then use the software\footnote{https://github.com/pennucci/PulsePortraiture} introduced in \citet{pdr14} along with 2D eight sub-band noise-free templates to generate TOAs. The telescope-dependent 2D templates were created by using the results of the frequency-dependent pulse profile evolution analysis (Perera et al. in preparation) which was based on the high-resolution observations reported in \citet{dlc+14}. We note that these TOAs improved the weighted rms of the timing residuals of LT observations alone by a factor of two compared to that obtained using the standard bandwidth-averaged TOA method \citep{tay92}.

The EFF and WSRT backends use bandwidths a factor of more than 2 smaller (see Table~\ref{info}) and, therefore,  the frequency-dependent profile variation becomes less significant. Thus, we use the standard technique \citep{tay92} for the remaining data.

\section{Timing the pulsar}
\label{timing}

We fit a previously published timing model \citep{dcl+16} to our observed TOAs and update the solution by minimising the chi-square of timing residuals using the pulsar timing software package TEMPO2 \citep{ehm06,hem06}. Since the pulsar is in a binary orbit, the timing model includes Keplerian and Post-Keplerian (PK) parameters \citep[see][]{lk05}. We combine TOAs from different telescopes/backends by fitting for time offsets or `JUMPs' in the timing model to account for any systemic delays between the data sets \citep[e.g.][]{vlh+16}. Note that some of our observations were obtained simultaneously using all/several telescopes as a part of LEAP observations \citep{bjk+16}. Therefore, these simultaneous observations further help in constraining the `JUMPs' between the relevant data sets in the timing analysis.
We note that the interferometric delay measurements between telescopes were not used for these LEAP observations to correct their relative time offsets. We determine the white noise of the pulsar by using the ``TEMPONEST''\footnote{https://github.com/LindleyLentati/TempoNest} \citep{lah+14}  plugin that is based on a Bayesian analysis. We include white noise parameters EFAC, $E_f$, and EQUAD, $E_q$, for each telescope/backend separately in our timing model, which are related to a TOA with uncertainty $\sigma_{t}$ in micro-seconds as $\sigma = \sqrt{E_q^2 + E_F^2 \sigma_{t}^2}$ \citep[see][]{lah+14,vlh+16}. We use uniform and log-uniform prior distributions for EFACs and EQUADs in the fit, respectively. We note that our parallax measurement is barely consistent with those reported in other studies \citep[see][]{zsd+15,dcl+16,vlh+16}. This could be due to different data lengths used in different analyses (e.g. 4.3 yrs of data in our work compared to 17.7 yrs of data in \citet{dcl+16}), or some covariance between parameters. Therefore, we use their parallax of $0.90(3)$ and keep it fixed.

\begin{figure}
\includegraphics[width=8.5cm]{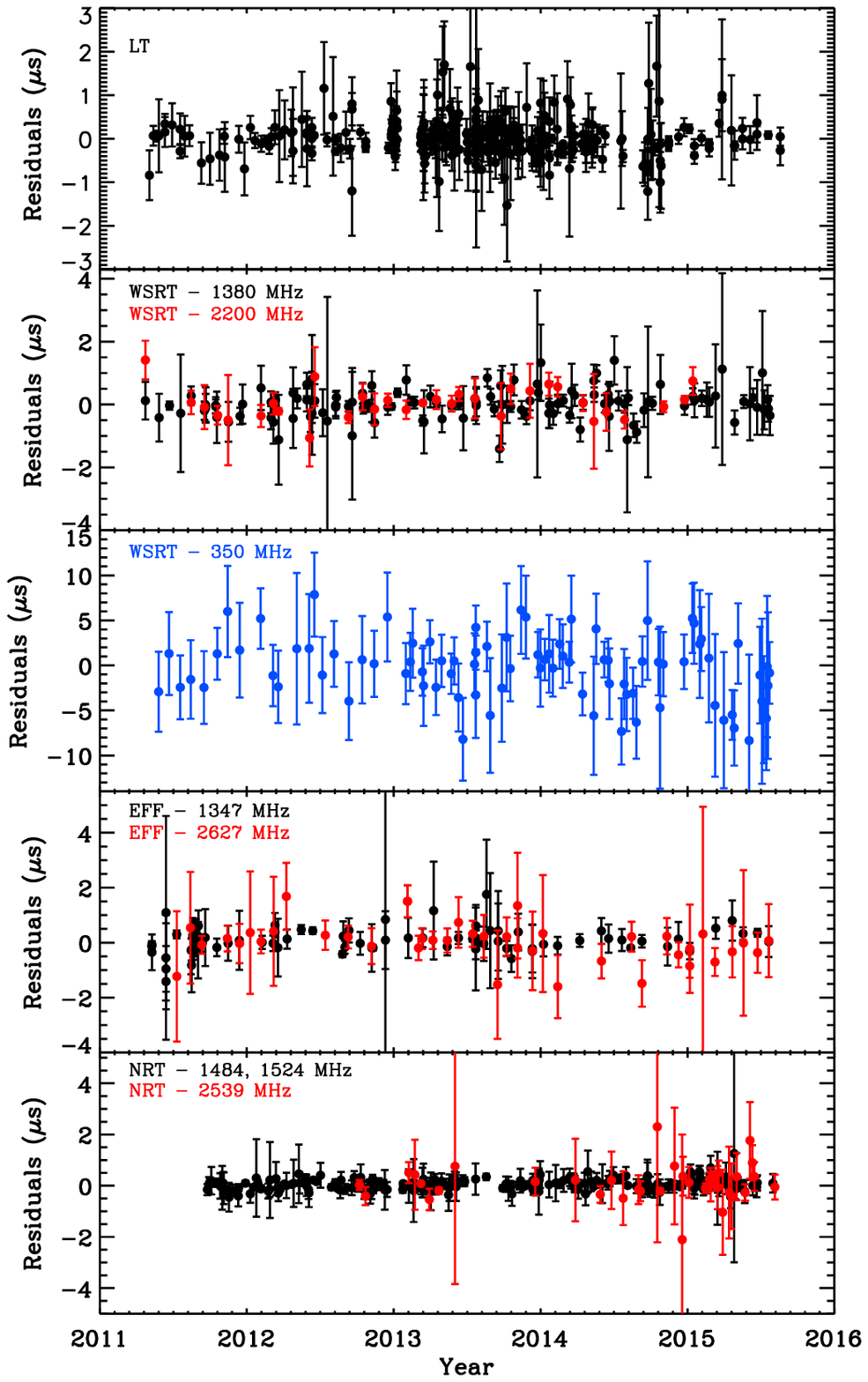}
\caption{
The timing residuals of PSR J1713$+$0747. The weighted rms of the residuals is 219~ns. For clarity of the plot, the residuals of the different telescopes are shown in separate panels. The $\sim$1.4~GHz observations are presented in {\it black}, and the low- and high-frequency observations are presented in {\it blue} and {\it red} colours, respectively. Note that each panel shows different scale in residuals. The average cadence of observations is 1.6 days across the entire data span, while the very-high-cadence observation period between 2013 February and 2014 April achieved an approximately daily observation cadence. Note that the WSRT 350~MHz data set has comparably large uncertainties and thus its weight in the timing analysis is small, although it contributes to the significance of the DM parameters.
}
\label{residuals}
\end{figure}

The observing frequencies of our data sets span $\sim$$350-2627$~MHz (see Table~\ref{info}). This broad frequency range allows us to fit for the ${\rm DM}$ and its first two time derivatives, $\rm \dot{DM} (\equiv dDM/dt)$ and $\rm \ddot{DM} (\equiv d^2DM/dt^2)$, in the timing model, and we measure a $\rm \ddot{DM}$ with a 3.5$\sigma$ significance. The pulsar showed a significant DM event towards the end of 2008 \citep[see][]{dcl+16,abb+15,zsd+15}. However, the TOAs in our data sets begin in 2011 and therefore, no effect from this event is seen in the analysis and we see no evidence for any subsequent events. The timing residuals shown in Figure~\ref{residuals} are based on the timing model given in Table~\ref{solution}. We note that these results were obtained without including red and stochastic DM noise modelling of the pulsar in the timing analysis.

\begin{table*}
\begin{center}
\caption{
The timing model parameters of PSR J1713$+$0747 constrained using our data set given in Table~\ref{info}. The binary parameters were measured using the T2 binary model given in TEMPO2. We keep the parallax fixed at the value $0.90(3)$ reported in \citet{dcl+16}. The position, spin frequency, and DM are given for the reference epoch of MJD 56000.
}
\label{solution}
\begin{tabular}{lcc}
\hline
\multicolumn{1}{l}{Timing parameter} &
\multicolumn{1}{c}{} & \\
\hline
Data span (MJD) & 55643 -- 57221 \\
Number of TOAs & 952 \\
Weighted rms timing residual ($\upmu$s) & 0.219 \\
Reduced $\chi^2$ value & 0.97 \smallskip \smallskip \\ 
Right ascension (RA) (J2000) & 17:13:49.5340822(7) \\
Declination (DEC)(J2000) & $+$07:47:37.48181(2) \\
Proper motion in RA (mas$/$yr) & 4.913(4) \\
Proper motion in DEC (mas$/$yr) & $-$3.955(8) \\
Spin frequency, $f$ ($\rm s^{-1}$) & 218.8118403818720(3) \\
Spin frequency $\rm 1^{st}$ time derivative, $\dot f$ ($\rm s^{-2}$) & $-4.08356(7)\times10^{-16}$ \\
Reference epoch (MJD) & 56000 \\
Parallax, $\pi$ (mas) & 0.9 \\
Dispersion measure, $\rm DM$ (cm$^{-3}$~pc) & 15.99191(4) \\
Dispersion measure $\rm 1^{st}$ time derivative, $\rm \dot{DM}$(cm$^{-3}$~pc~yr$^{-1}$) & $-1.0(2)\times10^{-4}$ \\
Dispersion measure $\rm 2^{nd}$ time derivative, $\rm \ddot{DM}$ (cm$^{-3}$~pc~yr$^{-2}$) & $2.8(8)\times10^{-5}$ \\
Orbital period, $P_{\rm b}$ (d) & 67.8251309805(6) \\
Epoch of periastron, $T_0$ (MJD) & 48741.9740(2) \\
Projected semi-major axis, $x$ (lt-s) & 32.34241993(5) \\
Longitude of periastron, $\omega$ (deg) & 176.201(1) \\
Orbital eccentricity, $e$ & $7.49404(6)\times10^{-5}$ \\
Orbital inclination, $i$ (deg) & 70.4(8) \\
Longitude of ascending node, $\Omega$ (deg) & 96(3) \\
Companion mass, $m_c$ (M$_\odot$) & 0.301(3) \smallskip \smallskip \\ 
Clock correction procedure & TT(BIPM2015) \\
Solar system ephemeris model & DE421 \\
Units & TCB \smallskip \smallskip \\
\hline
\end{tabular}
\end{center}
\end{table*}

The measurements of the two PK parameters, range $r$ and shape $s$, correspond to the Shapiro delay (i.e. the extra time delay of the signal due to the gravitational potential of the binary companion), which leads to measurements of the mass of the pulsar $m_{\rm p} = 1.41(6)$~M$_\odot$ and the companion $m_{\rm c} = 0.301(8)$~M$_\odot$ with a 2$\sigma$ uncertainty. We find that these measurements are consistent with the values reported in previous studies within their uncertainties \citep[see][]{zsd+15,dcl+16,vlh+16}.

PSR J1713$+$0747 exhibits some low-level red noise and stochastic DM noise in longer data sets \citep[e.g.][]{zsd+15,cll+16,lsc+16}.
The length of our data set is perhaps not long enough to perform a proper noise analysis of the pulsar to measure the appropriate parameters significantly. However, for comparison, we determine the red and DM stochastic noise parameters of the pulsar based on our data set using ``TEMPONEST'' \citep[see][]{lbj+14,lah+14}. We include a power-law DM to model a stochastic DM variation component in addition to $\rm \dot{DM}$ and $\rm \ddot{DM}$ in the timing model. To model an additional achromatic red noise process, we include a power-law red noise model and fit for all white and red noise terms along with the parameters given in the timing model simultaneously. We find that the timing measurements are consistent with those obtained without including these additional noise parameters in the timing model. We find the amplitude and the spectral index of the power-law DM variation are $\log A_{\rm DM}=-14.4 \pm 2.5$ and $\gamma_{\rm DM} = 2.3 \pm 1.8$, respectively, while those of the power-law red noise of the pulsar are $\log A_{\rm red}=-14.9 \pm 1.7$ and $\gamma_{\rm red} = 2.0 \pm 1.7$, respectively. Note that these noise parameters are poorly constrained due to our short data span, but consistent with those estimated in previous studies for this pulsar within the uncertainties \citep[see][]{dcl+16,vlh+16,cll+16}. We estimate the Bayes factor (i.e. $B=evidence[H_1]/evidence[H_0]$, where H$_1$ is the model including the red and DM stochastic noise and H$_0$ is the model excluding these noise terms) to be $\log_{10}({\rm B})=8.8(2)$, indicating that the noise model is preferred \citep{kr95}.

Note that we use the solar system ephemeris DE421 model \citep{fwb09} in our timing analysis (see Table~\ref{solution}). For an additional comparison, we use the more recent DE436\footnote{https://naif.jpl.nasa.gov/pub/naif/JUNO/kernels/spk/de436s.bsp.lbl} model in the timing analysis and confirm that the timing and noise parameters are very similar to those obtained with the DE421 model. We also confirm that the GW search results that obtained from the two models are similar and consistent with each other.

\section{Pulsar sensitivity to individual gravitational wave sources}
\label{sens}

The GW signal produced by a SMBHB can be seen in the pulsar timing residuals if the pulsar has a sufficiently precise timing solution and the  strain amplitude of the particular GW is large enough to produce a detectable signal in the timing residuals that can be distinguished from the timing noise. The sinusoidal GW signature in timing residuals produced by a SMBHB in a circular orbit has two terms due to its binary evolution, namely the `Earth term', which has a high-frequency behaviour, and the `pulsar term', which has a low-frequency behaviour \citep[see][]{jllw04}. We use the GW model given in \citet{bps+16}, and briefly describe the relevant expressions here in Appendix~\ref{theo}. In this study, we mainly assume non-evolving SMBHBs (i.e. the long-term evolution of the binary is negligible compared to the light travel time between the pulsar and the Earth), so that both the pulsar and Earth terms have the same frequency. We note that the sources might or might not be evolving at the high frequencies relevant to this study depending on the mass of the SMBHB. However, \citet{bps+16} showed that the GW upper limits are insensitive on the assumed source model (as we will also confirm this in Section~\ref{babak}), which indicates this working assumption is valid. By searching for the embedded GW periodic signal in timing residuals, we would be able to estimate the sensitivity limit of the pulsar on the strain amplitude of the CGWs produced by SMBHBs. We followed two different methods. First, we used computationally inexpensive spectral fitting method given in \citet{yhj+10}. We then undertook a computationally expensive more advanced Bayesian approach using the ``TEMPONEST'' plugin in TEMPO2. For an independent comparison, we finally use a previously developed algorithm for the first EPTA data release based on Bayesian analysis \citep{bps+16}, including evolving SMBHBs. We find both Bayesian analyses give consistent upper limits on the strain amplitude, while the spectral method does not provide optimal results, rather it provides slightly larger upper limits. As discussed in \citet{ejm12}, this spectral fitting method is incoherent and thus, provides non-optimal results. For completeness, we first discuss the spectral fitting method and then the Bayesian approaches.

\subsection{Spectral fitting method}
\label{spectral}

As mentioned before, here we use the spectral fitting method given in \citet{yhj+10}. The definition of the GW signal given in \citet{yhj+10} (see Equation 4 therein) represents a reduced form and thus, misses some parameters (e.g. orbital inclination $i$, initial phase $\phi_0$ of the GW signal) compared to the full expression given in Appendix~\ref{theo}. We notice that, this reduced form provides a few factors poor results in the upper limit estimates. Therefore, we use the full GW expression. 

\begin{figure}
\includegraphics[width=8cm]{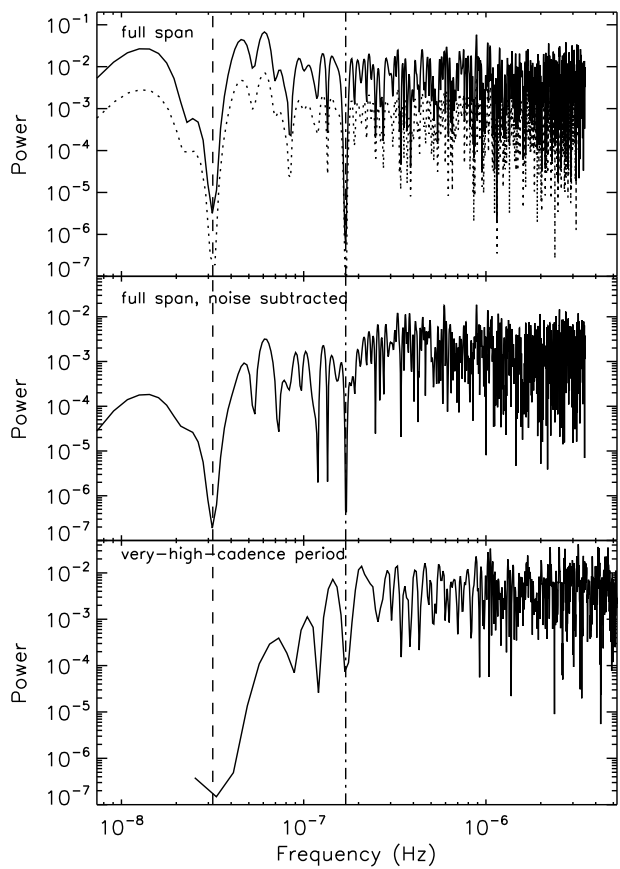}
\caption{
The power spectrum of the timing residuals of the pulsar given in Figure~\ref{residuals}. The {\it top} panel shows the spectrum generated using the entire data set across the full data span, including ({\it solid} curve) and excluding ({\it dotted} curve) the low frequency WSRT 350~MHz data set. For the clarity of the plot, we shifted the {\it dotted} curve down by a factor of 0.1. We note that both curves are very similar, indicating that the GLSP method accounts for the less weighted WSRT 350~MHz data correctly in the periodogram. The {\it middle} panel represents the power spectrum generated using the timing residuals after subtracting the wave forms of the red and stochastic DM power-law noise terms as described in Section~\ref{timing}. The {\it bottom} panel shows the spectrum for the very-high-cadence period of the observations after accounting for all noise parameters in the timing model. Note that the upper bound of the frequency range is extended beyond 3.5~$\mu$s with this very-high-cadence observations. The y-axis represents the normalised power \citep[see][for details]{zk09}. The vertical {\it dashed} and {\it dashed-dotted} lines represent the frequencies of 1~yr$^{-1}$ and the orbital period of 67.8~d of the pulsar, respectively.
}  
\label{psd}
\end{figure}

We first obtain the power spectrum of the timing residuals of PSR J1713$+$0747 shown in Figure~\ref{residuals}. The power spectrum of an unevenly sampled data set can be obtained by using the Lomb-Scargle periodogram \citep{lom76,sca82}. However, the original LS periodogram does not account for the uncertainties of the data. When generating the power spectrum from timing residuals, it is important to include the uncertainties to avoid any biases of the power in the spectrum to less-weighted residuals. Therefore, we use the Generalised Lomb-Scargle periodogram (GLSP) formalism introduced in \cite{zk09}, which accounts for the uncertainties of the data points by including weights and also fits for a floating-mean. The power spectrum of our timing residuals is given in Figure~\ref{psd}. We note that since the periodogram is weighted by the uncertainties, the low-frequency WSRT 350~MHz residuals with large uncertainties (see Figure~\ref{residuals}) do not contribute significantly. This can be clearly seen in the {\it top} panel  in Figure~\ref{psd}, where the spectrum is calculated including ({\it solid} curve) and excluding ({\it dotted} curve) the WSRT 350~MHz data, respectively. Note that we shifted the {\it dotted} curve down by a factor of 0.1 for clarity of the figure.

In contrast to PSR B1937$+$21 \citep[see][]{yss+14}, Figure~\ref{psd} shows that the timing residuals of PSR J1713$+$0747 do not present any significant unmodelled periodic signals. However, as mentioned in Section~\ref{timing}, we also estimated DM stochastic and red noise parameters in the timing analysis. For comparison, the middle panel in Figure~\ref{psd} shows the power spectrum after subtracting the time-domain waveforms of these noise terms in the timing model, resulting in a slightly lower spectral power at low-frequencies compared to the previous case. The power spectrum of the residuals, after subtracting the wave forms of additional DM and red noise terms, on the data within the very high-cadence observing period is shown in the {\it bottom} panel of Figure~\ref{psd}. No noticeable red noise is seen within this short, but high-cadence,  time span. The significant power drop at the yearly period (i.e. a frequency of $\sim 3\times10^{-8}$~Hz) in Figure~\ref{psd} is common for all three cases and it is due to fitting for the pulsar position in the timing model. Similarly, the power absorption at $\sim 1.7\times10^{-7}$~Hz is due to fitting for the orbital period, $P_{\rm b} = 67.8$~d.

We then use the same method outlined in previous studies \citep[see][]{yhj+10,yss+14} to build the detection threshold using the power spectrum of timing residuals. To remove any spurious effects from any apparent unmodelled signals and noise in the timing residuals, we smooth the spectrum given in Figure~\ref{psd} (top panel) and fit a polynomial. We find that a third-order polynomial function is sufficient to fit the data. We then scale this polynomial by a factor $\alpha$ in power in the spectrum. It is determined  by simulating $10^4$ data sets with TOAs having uncertainties of 100~ns and cadence that matches the observed data. We then obtain the power spectrum of each realisation and calculate the mean power. We label the mean power of the $i$th spectrum as $m_{\rm i}$. We increase $\alpha$ starting from 1 and count the number of power spectra that have any power value greater than $\alpha m_{\rm i}$. We set a 1\% false alarm rate and increase $\alpha$ until the number of spectra satisfies this threshold. We find that the best $\alpha$ for our data set is 10.1, and scale the polynomial in power accordingly, in order to build the detection threshold.

\begin{figure}
\includegraphics[width=7.5cm]{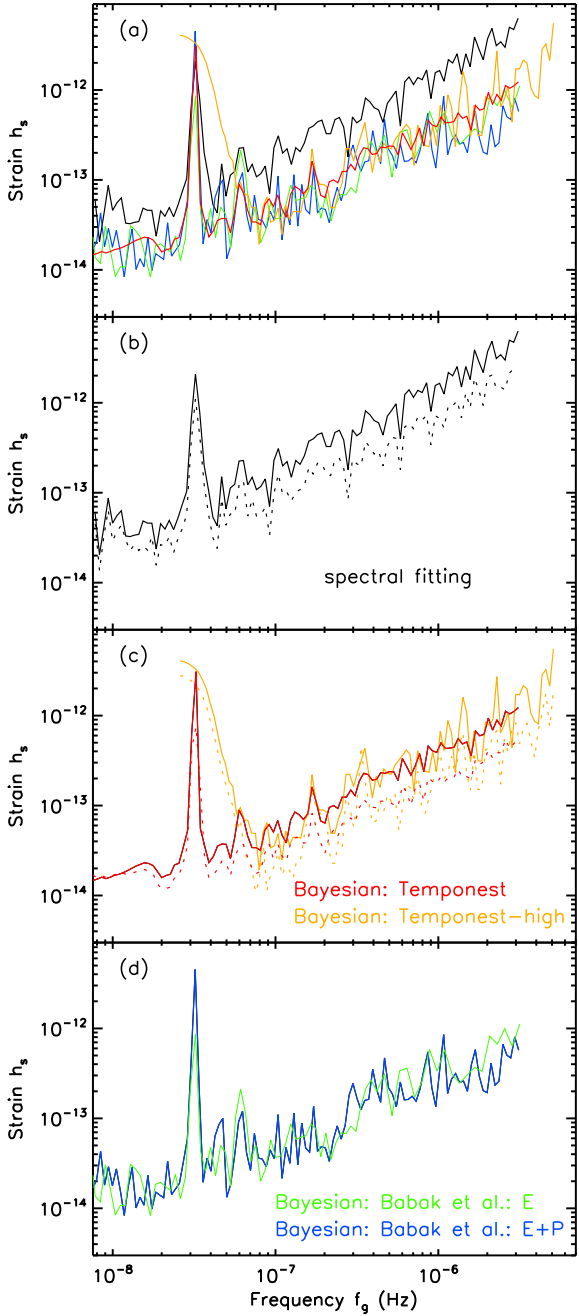}
\caption{
The  95 per cent upper limit on the GW strain amplitude produced by SMBHBs based on the timing observations of PSR J1713$+$0747. Panel (a) shows the results obtained from all methods described in Section~\ref{sens}. For clarity, we plot the results obtained from each method in the following panels (i.e. from spectral fitting, and Bayesian analysis using Temponest and the algorithm given in \citet{bps+16}, respectively). The {\it dotted} curves in panel (b) and (c) show the upper limits for the optimal case where all the sources are assumed to be located in the same direction as the pulsar. For comparison, we over plot the upper limits obtained from the data within the high-cadence observation time span between February 2013 and April 2014 ({\it orange} curves in panel (c)). Panel (d) shows the upper limits obtained from the algorithm given in \citet{bps+16} assuming evolving sources, for only the Earth term ({\it green}) and also for the combined Earth and pulsar terms ({\it blue}). 
}
\label{result1}
\end{figure}

We then divide the frequency range of the observed spectrum into 100 equal bins in log-space. We inject the pulsar and Earth terms at the same frequency, according to the model given in Appendix~\ref{theo}, for a given $h_s$ in the observed TOAs (see Equation~\ref{signal}). We fit our timing model of the pulsar to the new GW-injected TOAs using TEMPO2 and obtain timing residuals. The power spectrum of the timing residuals is then compared with the detection threshold at the given frequency for a detection \citep[see][for more details]{yhj+10,yss+14}. For each $h_s$, we perform 1000 trials using randomly selected $\phi_S$, $\theta_S$, $\phi_0$, $\psi$, and $i$ and then count the number of detections (see Appendix~\ref{theo} for definitions of these parameters). We increase $h_s$ until the signals of 95 per cent of trials are detected and then record the particular $h_s$ as the upper limit for that given frequency bin. We repeat this process for all frequency bins. This gives the 95 per cent upper limit on the sky-averaged strain amplitude of CGWs produced by SMBHBs. We present our upper limit curve in Figure~\ref{result1} (see panel (a) and (b)). For an optimal sky location, we assume all the SMBHBs are located in the direction of the pulsar ($\pm10^\circ$ in the direction of the pulsar), and repeat the same procedure as before. As shown in the figure, the optimal upper limit is a factor of a few better than the sky-averaged result.

\subsection{Bayesian approach using TEMPONEST}
\label{bayes}

The spectral fitting method described above is an incoherent ad-hoc method that does not provide optimal results. It does not include the noise processes of the pulsar (i.e. red and DM stochastic noise) and it is not straightforward to implement them in the model. If the noise modelling is included in the analysis, then we need to fit for noise properties each time the GW signal is injected and thus, the fitting method becomes computationally inefficient. Therefore, we use the Bayesian approach described in \citet{lah+14} where the noise properties of the pulsar are constrained using the ``TEMPONEST'' plugin. Since we assume that the GW is produced by a SMBHB in a circular orbit, the signal that can be embedded in the data of a single pulsar is sinusoidal. Therefore, we fit for an additional simple sinusoidal signal while fitting for white noise (i.e. EFACs and EQUADs), and power-law stochastic DM and  red noise parameters of the pulsar simultaneously using ``TEMPONEST''. The timing model parameters are marginalised over during the fit  \citep[see][for details about this procedure]{lah+14}. The plugin uses Bayesian inference tool MULTINEST to determine this joint parameter space and nested Markov Chain Monte Carlo method for sampling \citep{fh08,fhb09,fhcp13}. We use log-uniform prior distributions for amplitudes of the power-law DM stochastic and red noise, and for the frequency of the sinusoidal signal. We use uniform prior distributions for spectral indices of the power-law noise terms, and the amplitude and the phase of the sinusoidal signal. The fit results in approximately $4\times10^4$ samples in the posterior probability distribution. Each sinusoidal signal sample includes information about its amplitude (r(t)), initial phase ($\phi_0$), and the frequency ($f_g$). We then convert the amplitudes of these GW signals to strain amplitudes ($h_s$) using Equation~\ref{signal} for randomly selected $\phi_S$, $\theta_S$, $\phi_0$, $\psi$, and $i$ (i.e. the polar coordinates of the GW source, initial phase of the signal, GW polarisation angle, and orbital inclination, respectively -- see Appendix~\ref{theo} for details). We divide the frequency range of these GWs (which is equal to the frequency range used in Section~\ref{spectral}) into 100 equal bins and then obtain the 95 per cent upper limit on the strain amplitude of corresponding signals in each bin. Figure~\ref{result1} shows our results (see panel (a) and (c)). The optimal upper limit is again obtained by considering all sources to be located along the direction of the pulsar (see panel (c) in Figure~\ref{result1}) as described in Section~\ref{spectral}.

As mentioned above, Figure~\ref{result1} shows that the Bayesian approach provides better limits compared to the spectral fitting method. We also applied this method to the data within the high-cadence observation time span and obtained the upper limit on the strain amplitude  (panel (c) of Figure~\ref{result1}).

\subsection{Using previously published Bayesian algorithm}
\label{babak}

In order to further validate our results, we use the algorithm presented in \citet{bps+16}. In this Bayesian analysis, we use PolyChord \citep{hhl15} to fit for the CGW signal (i.e. $h_s$, $\phi_S$, $\theta_S$, $\phi_0$, $\psi$, $i$, and $f_g$) and the noise parameters (i.e. EFACs, EQUADs, and power-law stochastic DM, and red noise processes). Similar to the method in Section~\ref{bayes}, the timing model parameters are marginalised over during the fit. In addition to different tools used in the two Bayesian approaches, \citet{bps+16} is capable of using both evolving and non-evolving sources in the search analysis, whereas the method in Section~\ref{bayes} uses only non-evolving sources. The underlying algorithm utilises the  GW model described in Appendix~\ref{theo}, and the details of the fitting procedure is described in \citet{bps+16} and \citet{nx01}. In the search for evolving sources, we use three extra parameters in the model -- the chirp mass of the system, distance to the pulsar, and the initial phase of the pulsar term. For the distance to the pulsar we use the Gaussian prior centred at the best currently known value with the measured uncertainty of  1.05(7)~kpc \citep{cbv+09}. As mentioned above, we use PolyChord (nested sampling) in this analysis, which is the next incarnation of the MULTINEST. It is supposed to be more robust for the multi-modal likelihood surfaces embedded in the large dimensional parameter space and more efficient for some problems. However, we find no benefit of the new implementation, rather both samplers give very consistent results. The posterior distribution for the noise parameters obtained by the TEMPONEST (described in the Section~\ref{bayes}) and the method used here are very similar. We also note that the results obtained from evolving and non-evolving sources are consistent with each other. We present the upper limit on the strain amplitude of CGWs in Figure~\ref{result1} (see panel (a) and (d)) separately for the Earth term only, as well as for the combined Earth and pulsar terms. Figure~\ref{result1} shows that these results are consistent with the results obtained from the Bayesian approach described in Section~\ref{bayes}.

\subsection{Application to proposed SMBHB candidates}
\label{application}

\begin{figure}
\includegraphics[width=8cm]{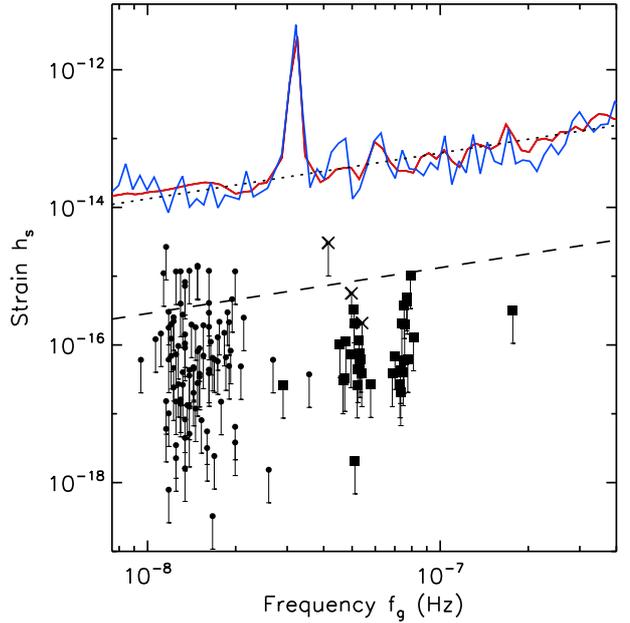}
\caption{
The 95 per cent upper limit on the sky-averaged strain amplitude of CGWs. The colour code of the curves is the same as that given in Figure~\ref{result1}. For clarity, only the limits obtained from Bayesian approaches are plotted. The {\it dashed} and {\it dotted} lines show the theoretically expected strain amplitude from SMBHBs with equal masses of $10^{9}$~M$_\odot$ and $10^{8}$~M$_\odot$ at a distance of the Virgo cluster, 16.5~Mpc, respectively. The {\it circles}, {\it squares}, and {\it crosses} show the expected GW strain amplitude produced by the SMBHB candidates reported in \citet{gds+15}, \citet{cbh+16}, and \citet{lgb+16}, respectively (see Table~\ref{candidates} for more information). Note that the expected strain amplitudes are estimated using Equation~\ref{expect} given in Appendix~\ref{theo}. 
}
\label{sensit}
\end{figure}

It is thought that quasars contain SMBHs and SMBHBs in their nuclei. Although SMBHBs are not resolvable optically by direct imaging, 
candidates can be identified through periodicity in their optical, radio, and X-ray fluxes. The quadrupole torque produced by a binary induces a periodicity in the accretion flow from a circumbinary disk related to its orbital period \citep{al94,mm08,rds+11}. This might result in periodic luminosity variations \citep[e.g.][]{srr+12}. By searching for periodicities in observations of quasars, SMBHB candidates have been identified in many surveys. In this study we consider those identified by the Catalina Real-time Transient Survey (CRTS) \citep{gds+15}, Palomar Transient Factory (PTF) data base \citep{cbh+16}, and Pan-STARRS1 Medium Deep Survey \citep{lgh+15,lgb+16}, which identified 111, 33, and 3 candidate SMBHBs, respectively (see Table~\ref{candidates}). We note that the periodicities of all these candidate sources are expected to produce GWs in the frequency range between approximately  $10^{-8}-10^{-7}$~Hz, which falls into the frequency range that our timing of PSR J1713$+$0747 can probe. Therefore, we estimated these expected signals (using Equation~\ref{expect} and assuming equal mass SMBHs (i.e. $q=M_2/M_1=1$) in binary systems) and compared them with our timing derived upper limit on the strain amplitude (see Figure~\ref{sensit}). The lower limit of these expected signals shown in the figure are estimated by assuming a mass ratio of $q=M_2/M_1= 0.1$. As seen in Figure~\ref{sensit}, the strain amplitudes of the predicted GW signals from these sources are lower (by more than a factor of 4 for the strongest ones) than the sky-averaged sensitivity curve for PSR J1713$+$0747. We note that none of these candidates are located in the direction of the pulsar. Thus, using the actual sky locations of these sources in the GW search analysis will not improve the sensitivity significantly compared to the sky-averaged sensitivity presented in Figure~\ref{sensit} (see the dotted curves in Figure~\ref{result1} for upper limits based on optimal sky location). This indicates that, we cannot expect to detect GW signals produced by these candidates in our timing results yet.

\begin{table}
\begin{center}
\caption{
The SMBHB candidates considered in this analysis based on three studies; (1) Catalina Real-time Transient Survey (CRTS) \citep{gds+15}, (2) Palomar Transient Factory (PTF) data base \citep{cbh+16}, and (3) Pan-STARRS1 Medium Deep Survey \citep{lgh+15,lgb+16}. Note that the expected GW frequencies from these sources fall in the regime that our pulsar timing analysis can be probed.  
}
\label{candidates}
\begin{tabular}{ccccc}
\hline
\multicolumn{1}{c}{Study} &
\multicolumn{1}{c}{Candidates} &
\multicolumn{1}{c}{$z$} &
\multicolumn{1}{c}{Period} &
\multicolumn{1}{c}{Frequency} \\
\multicolumn{1}{c}{ } &
\multicolumn{1}{c}{ } &
\multicolumn{1}{c}{} &
\multicolumn{1}{c}{(yr)} &
\multicolumn{1}{c}{(nHz)} \\
\hline
1 & 111 & $0.1-2.7$ & $1.8-6.7$ & $9.5-35.7$ \\
2 & 33 & $0.3-3.1$ & $0.3-2.1$ & $29-177$ \\
3 & 3 & $1.2-2.2$ & $1.2-1.5$ & $41-54$\\
\hline
\end{tabular}
\end{center}
\end{table}

\section{Conclusions}
\label{dis}

We used high-cadence observations of PSR J1713$+$0747 to place upper limits on the strain amplitude of CGWs produced by individual SMBHBs in circular orbits. 
Based on the typical frequency range used in previous studies, $1/T_{\rm obs}$ and $N/(2T_{\rm obs})$ \citep[e.g.][]{yhj+10,yss+14}, we used our observations to probe GWs produced within a frequency range between $7.8\times10^{-9}$~Hz and $5\times10^{-6}$~Hz, covering the high-frequency $\upmu$Hz regime.
We used three independent methods in the analysis, including a spectral fitting method and more advanced Bayesian approaches. As mentioned in previous studies \citep[see][]{yhj+10,ejm12} and also shown in Figure~\ref{result1}, the spectral fitting method does not provide optimal results due to its simple incoherent fitting procedure and absence of proper noise modelling of the pulsar. We find that the independent Bayesian analyses are consistent, and also about a factor of five better compared to the spectral fitting method. Based on our results, we found a 95 per cent upper limit on the sky-averaged strain amplitude of CGWs to be $\lesssim3.5\times10^{-13}$ at a reference frequency of $1$~$\upmu$Hz. For an optimal sky location, the 95 per cent upper limit on the strain amplitude is improved to be $\lesssim2.1\times10^{-13}$ at the same reference frequency. We also found that our timing results place upper limits on the sky-averaged and optimal strain amplitude of low frequency CGWs to be $\lesssim 1.4\times10^{-14}$ and $\lesssim1.1\times10^{-14}$ at a reference frequency of $20$~nHz, respectively. This low-frequency limit is approximately a three orders of magnitude better compared to the result presented in \citet{yhj+10} for this pulsar. We also compared our limits with the expected GWs produced by SMBHB candidates, and found that the limits are not yet constraining these sources. However, we will be able to confirm or reject the binary hypothesis for those candidates in the future with better limits including more observations.

Compared to the upper limits of PSR B1937$+$21 on GWs presented in \citet{yss+14}, our study shows that PSR J1713$+$0747 provides better sky-averaged upper limits across the entire frequency range. The poor limits of PSR B1937$+$21 may have been due to excess noise caused by the known significant DM variation and the high-level of red noise in the timing data \citep[see][]{yss+14,cll+16,lsc+16}. As mentioned in Section~\ref{spectral}, we also note that \citet{yss+14} used a simplified GW model and it worsens the upper limits by factor of a few compared to that obtained from the full definition of the GW model that we used in our study. The impact of the presence of low-frequency noise in the timing data on their sensitivity to the strain amplitude of GWs was investigated by \citet{cll+16} using a subset of the EPTA data over a longer time span of about 17 years. They showed that the sensitivity reduces by a factor of up to $\sim$5 at nHz frequencies while at the higher frequencies that our  data set is most sensitive, the impact of noise is minimal.

In comparison, the sensitivity limits on CGWs estimated in our analysis at low-frequencies are consistent in general with those estimated by PTA studies \citep[see][]{bps+16,abb+14,zhw+14}. These PTA studies combined data from several pulsars in their analysis over long-term observations to obtain their sensitivities compared to single-source data over a short-term observations in our analysis. Our study indicates the importance of collecting more high-cadence observations from good pulsars in PTAs to improve the timing precision and then obtain better GW limits even at low-frequencies.

The precision of the timing model (improving the TOA uncertainty) can be improved with time by extending the timing baseline, using larger telescopes and modern backends to observe the pulsar \citep[see][]{lk05}. Therefore, in the future, we can improve the timing precision by including more data in our analysis and thus, improve the sensitivity limits of pulsar timing to GWs produced by SMBHBs.

\section*{Acknowledgments}
The European Pulsar Timing Array (EPTA: www.epta.eu.org) is a collaboration of European institutes to work towards the direct detection of low-frequency gravitational waves and the running of the Large European Array for Pulsars (LEAP). Pulsar research at the Jodrell Bank Centre for Astrophysics and the observations using the Lovell Telescope are supported by a consolidated grant from the STFC in the UK. The Nan\c{c}ay radio observatory is operated by the Paris Observatory, associated with the French Centre National de la Recherche Scientifique (CNRS). We acknowledge financial support from `Programme National de Cosmologie et Galaxies' (PNCG) of CNRS/INSU, France. The Westerbork Synthesis Radio Telescope is operated by the Netherlands Institute for Radio Astronomy (ASTRON) with support from The Netherlands Foundation for Scientific Research NWO. Part of this work is based on observations with the 100-m telescope of the Max-Planck-Institut f$\ddot u$r Radioastronomie (MPIfR) at Effelsberg. The LEAP is supported by the ERC Advanced Grant `LEAP', Grant Agreement Number 227947 (PI M. Kramer).

KL acknowledge the financial support by the European Research Council for the ERC Synergy Grant BlackHoleCam under contract no. 610058.
SO acknowledges support from the Alexander von Humboldt Foundation and Australian Research Council grant Laureate Fellowship FL150100148.
AS is supported by the Royal Society. GJ and GS acknowledge support from the Netherlands Organisation for Scientific Research NWO (TOP2.614.001.602). SRT is supported by the NANOGrav project, which receives support from National Science Foundation Physics Frontier Center award number 1430284. We thank Chiara Mingarelli and Kejia Lee for useful comments on the manuscript.

\bibliography{psrrefs,modrefs,journals}
\bibliographystyle{mnras}

\appendix
\section{GW model}
\label{theo}

The sinusoidal signal produced by a non-evolving GW source in the timing residuals of a pulsar is given by

\begin{equation}
\label{signal}
r(t) = r^{e}(t) - r^{p}(t_p).
\end{equation}

\noindent
Here, $t_p =  t - D(1 + \hat{\Omega} \cdot \hat{p})/c$, where $D$ is the distance to the pulsar, $c$ is the speed of light, $\hat{\Omega}$ is the unit vector along the direction of GW propagation, and $\hat{p}$ is the unit vector along the direction of the radio waves propagating from the pulsar.  
We use the astrometry-derived pulsar distance of $1.05(7)$~kpc \citep{cbv+09}, which is consistent with the distance derived from the measured DM and the parallax \citep[see][]{dcl+16}. The Earth and pulsar terms are given by

\begin{equation}
\begin{split}
r^{e}(t) = & \frac{h_s}{\omega} \{ (1+\cos^2i)F^{+} [\sin(\omega t + \phi_0)]  \\  & + 2\cos iF^{\times}[\cos(\omega t + \phi_0)] \}
\end{split}
\end{equation}

\noindent
and

\begin{equation}
\begin{split}
r^p(t_p) = & \frac{h_s}{\omega} \{ (1+\cos^2i)F^{+} [\sin(\omega t_p + \phi_0)]  \\  & + 2\cos iF^{\times}[\cos(\omega t_p + \phi_0)] \}
\end{split}
\end{equation}

\noindent
where, $h_s$ is the strain amplitude of the GW signal, $f_g = \omega/2\pi$ is the frequency of the GW signal, $i$ is the inclination angle of the SMBHB orbit with respect to the line-of-sight, $\phi_0$ is the initial phase of the signal. The ``antenna beam patterns'' are given by

\begin{equation}
F^{+} = \frac{1}{2} \frac{(\hat{m}\cdot\hat{p})^2  -  (\hat{n}\cdot\hat{p})^2}{1 + \hat{\Omega}\cdot\hat{p}}
\end{equation}

\noindent
and

\begin{equation}
F^{\times} =  \frac{(\hat{m}\cdot\hat{p}) (\hat{n}\cdot\hat{p})}{1 + \hat{\Omega}\cdot\hat{p}}.
\end{equation}

\noindent
For a cartesian coordinate system (x, y, z), we can define all above unit vectors as

\begin{equation}
\begin{split}
\hat{m} = & (\sin\phi_S \cos \psi - \sin \psi \cos \phi_S \cos \theta_S)\hat{x} \\ & - (\cos \phi_S \cos \psi + \sin \psi \sin \phi_S \cos \theta_S)\hat{y} + (\sin \psi \sin \theta_S)\hat{z}
\end{split}
\end{equation}

\begin{equation}
\begin{split}
\hat{n} = & (-\sin\phi_S \sin \psi - \cos \psi \cos \phi_S \cos \theta_S)\hat{x} \\ & + (\cos \phi_S \sin \psi - \cos \psi \sin \phi_S \cos \theta_S)\hat{y} + (\cos \psi \sin \theta_S)\hat{z}
\end{split}
\end{equation}

\begin{equation}
\hat{\Omega} = -(\sin \theta_S \cos \phi_S)\hat{x} - (\sin \theta_S \sin \phi_S)\hat{y} - \cos\theta_S \hat{z}
\end{equation}

\begin{equation}
\hat{p} = (\sin \theta \cos \phi)\hat{x} + (\sin \theta \sin \phi)\hat{y} + \cos\theta \hat{z}
\end{equation}

\noindent
where, $\theta_S$ and $\phi_S$ are the usual polar coordinates of the GW source location in the sky, $\theta$ and $\phi$ are the usual polar coordinates of the pulsar, and $\psi$ is the GW polarisation angle. The derivation of the above expressions are given in previous studies in detail \citep[see][]{ljp08,sv10, lwk+11, zhw+14, bps+16}. The GW strain amplitude is defined as 

\begin{equation} 
\label{expect}
h_s = 2 \frac{(GM_c)^{5/3}}{c^4 D_L(z)} (\pi f_g)^{2/3}
\end{equation}

\noindent
where, $M_c = (M_1M_2)^{3/5}(M_1+M_2)^{-1/5}$ is the chirp mass of the system with individual masses of $M_1$ and $M_2$, and $G$ is the gravitational constant. The luminosity distance $D_L$ to the source is given by

\begin{equation} 
D_L(z) = (1+z)\frac{c}{H_0} \int_0^z \frac{dz'}{E(z')},
\end{equation}

\noindent
where, $z$ is the redshift to the source, $H_0$ is Hubble's constant ($=$72~km/s/Mpc), and $E(z)=H(z)/H_0 = \sqrt{\Omega_\Lambda + \Omega_m(1+z)^3}$ \citep[see][]{saw09}. We use $\Omega_\Lambda=0.7$ and $\Omega_m=0.3$ in this work \citep[see][]{cp05,kdn+09}.

\end{document}